\documentstyle[12pt]{article}

\begin{document}

\renewcommand{\baselinestretch}{1.5}


\catcode`\@=11
\renewcommand{\theequation}{\thesection.\arabic{equation}}
\@addtoreset{equation}{section}
\catcode`\@=10

\def\dsp{\displaystyle}
\def\d{\mbox{\rm d}}
\def\e{\mbox{\rm e}}


\def\ut#1{{\mathunderaccent\tilde #1}}
\def\cosec{{\rm cosec}}
\def\kd{{\delta _{ij}}}
\def\ke{{\epsilon _{ijk}}}
\def\ha{\mbox{$\frac{1}{2}$}}
\def\tha{\mbox{$\frac{3}{2}$}}
\def\fha{\mbox{$\frac{5}{2}$}}
\def\nha{\mbox{$\frac{n}{2}$}}
\def\oei{\mbox{$\frac{1}{8}$}}
\def\oqr{\mbox{$\frac{1}{4}$}}
\def\tqr{\mbox{$\frac{3}{4}$}}
\def\otw{\mbox{$\frac{1}{12}$}}
\def\oth{\mbox{$\frac{1}{3}$}}
\def\ofi{\mbox{$\frac{1}{5}$}}
\def\osi{\mbox{$\frac{1}{6}$}}
\def\fei{\mbox{$\frac{5}{8}$}}
\def\fra#1#2{\mbox{$\frac{#1}{#2}$}}
\def\f{\frac}
\def\p{\dsp\partial}
\def\arcsinh{{\rm arcsinh}}
\def\sinh{{\rm sinh}}
\def\cosh{{\rm cosh}}
\def\arccosh{{\rm arccosh}}
\def\arctanh{{\rm arctanh}}
\def\tanh{{\rm tanh}}
\def\coth{{\rm coth}}
\def\sech{{\rm sech}}
\def\cosech{{\rm cosech}}


\def\p{\dsp \partial}

\def\pb{{\p\over\p b}}
\def\pc{{\p\over\p c}}
\def\pd{{\p\over\p d}}
\def\pe{{\p\over\p e}}
\def\pf{{\p\over\p f}}
\def\pg{{\p\over\p g}}
\def\ph{{\p\over\p h}}
\def\pj{{\p\over\p j}}
\def\pk{{\p\over\p k}}
\def\pl{{\p\over\p l}}
\def\pn{{\p\over\p n}}
\def\po{{\p\over\p o}}
\def\pq{{\p\over\p q}}
\def\pr{{\p\over\p r}}
\def\ps{{\p\over\p s}}
\def\pt{{\p\over\p t}}
\def\pu{{\p\over\p u}}
\def\pv{{\p\over\p v}}
\def\pw{{\p\over\p w}}
\def\px{{\p\over\p x}}
\def\py{{\p\over\p y}}
\def\pz{{\p\over\p z}}


\def\pa#1#2{\f{\dsp \p#1}{\dsp \p#2}}
\def\pp#1#2{\f{\dsp \p^2 #1}{\dsp \p #2^2}}
\def\pam#1#2#3{\f{\dsp \p^2 #1}{\dsp \p #2 \p #3}}

\def\dm#1#2{\f{\dsp \d #1}{\dsp \d #2}}
\def\dd#1#2{\f{\dsp \d^2 #1}{\dsp \d #2^2}}

\def\dddot#1{\mathinner{\buildrel\vbox{\kern5pt\hbox{...}}\over{#1}}}


\def\be{\begin{equation}}
\def\ee{\end{equation}}
\def\bq{\begin{eqnarray}}
\def\eq{\end{eqnarray}}
\def\beq{\begin{eqnarray*}}
\def\eeq{\end{eqnarray*}}
\def\ext#1#2{G^{[#1]} #2_{|_{\!\!|_{#2=0}}} = 0}
\def\exta#1{G^{[#1]}}
\def\extb#1#2{G^{[#1]} #2_{|_{\!\!|_{#2=0}}} }
\def\enter{$\longleftarrow\!\!\!^|$}
\def\abstract#1{\begin{minipage}{350pt}{\large\bf Abstract}: #1 \end{minipage}}
\def\heading#1{\begin{center} {\LARGE #1} \end{center} \vspace{10mm}}

\def\de{differential equation}
\def\des{differential equations}
\def\fode{first order ordinary differential equation}
\def\fodes{first order ordinary differential equations}
\def\pde{partial differential equation}
\def\pdes{partial differential equations}
\def\sodes{second order ordinary differential equations}
\def\Sodes{Second order ordinary differential equations}
\def\sode{second order ordinary differential equation}
\def\nodes{$n$th order ordinary differential equations}
\def\pain{Painlev\'{e}}
\def\rps{Right \pain\ series}
\def\lps{Left \pain\ series}
\def\lob{leading order behaviour}


\def\orl{Orl\'{e}ans}
\def\ie{{\it ie }}
\def\cf{{\it cf }}
\def\viz{{\it viz }}
\def\etal{{\it et al }}
\def\n{\nonumber}
\def\({\left (}
\def\){\right )}
\def\bi{\begin{itemize}}
\def\ei{\end{itemize}}
\def\z{&=&}
\def\lb{\left[ }
\def\rb{\right] }

\def\re#1{(\ref{#1})}

\def\article1#1#2#3#4{

 \titlepage
 \begin{center}
 \vspace*{20mm}
 {\huge \bf #1}

 \vspace{15mm}

 {\Large #2}

 \vspace{10mm}

 {#3}

 \end{center}

 \begin{center}
 \abstract{#4}

 \end{center}
}

\article1{Symmetry, singularities and integrability in complex dynamics II:
Rescaling and time-translation in two-dimensional systems}
{P G L Leach\footnote{Permanent address:  School of
Mathematical and Statistical Sciences, University of Natal, Durban, South
Africa 4041}, S Cotsakis and G P Flessas}{GEODYSYC\\
Department of Mathematics\\
University
of the Aegean, Karlovassi 83 200, Greece}{The explicit integrability of second
order ordinary differential equations invariant under time-translation and
rescaling is investigated.  Quadratic systems generated from the
linearisable version of this class of equations are analysed to determine the
relationship between the Painlev\'e and singularity properties of the different
systems.  The transformation contains a parameter and for critical values,
intimately related to the possession of the Painlev\'e Property in the parent
second order
equation, one finds a difference from the generic behaviour.  This study is a
prelude to a full discussion of the class of transformations which preserve the
Painlev\'e Property in the construction of quadratic systems from scalar
\nodes\ invariant under time translation and rescaling.}

\section{\Sodes\ invariant under time-translation and rescaling}

The general form of the second order ordinary differential equation invariant
under the two symmetries representing invariance under time translation and
rescaling, \viz
\begin{equation} G_1 = \pt \quad G_2 = -t\pt+x\px \label{0.1}
\end{equation}
is
\begin{equation}
\frac{\ddot{x}}{x^3} + f\left(\frac{\dot{x}}{x^2}\right) = 0, \label{0.2}
\end{equation}
where $f $ is an
arbitrary function of its argument.  As a second order equation with two
symmetries \re{0.2} can always be reduced to quadratures.  In the particular
case that $f $ is linear with a specific form, \ie
\be
f = 3b\(\frac{\dot{x}}{x^2}\)+b^2, \label{0.2a}
\ee
giving the equation
\begin{equation}
\ddot{x} + 3bx\dot{x} + b^2x^3 = 0 \label{0.3}
\end{equation}
the equation possesses eight
Lie point symmetries, naturally with the algebra $sl (3,R) $, and so is
linearisable \cite{Mohammed1}.

Not only does \re{0.3} possess the Painlev\'e
property but it has both Left and Right Painlev\'e Series \cite{Leach}.  The
equation is a representative of the Riccati hierarchy \cite{Kamke}[6.39, p
550] and is integrable via transformation to a linear third order equation
using a generalised Riccati transformation.  We write
\begin{equation}
x = \alpha\frac{\dot{w}}{w}, \label{0.4}{}
\end{equation}
where $\alpha $ is a
constant to be determined.  Equation \re{0.3} becomes
\begin{equation}
\frac{\dddot{w}}{w} + 3\left(\alpha b - 1\right )\frac{\dot{w}\ddot{w}}{w^2}
+\left(b^2\alpha^2 - 3b\alpha + 2\right)\frac{\dot{w}^3}{w^3} = 0.
\label{0.5}
\end{equation}
Equation \re{0.5} is a third order linear
differential equation in the case that both
\begin{equation}
\alpha b - 1 = 0 \quad\mbox{\rm and}\quad b^2\alpha^2 - 3b\alpha + 2 = 0, \label{0.6}
\end{equation}
\ie $\alpha b = 1 $.  The solution is
\begin{equation}
w = A + 2Bt + Ct^2, \label{0.7}
\end{equation}
where $A $, $B $ and $C $ are arbitrary
constants.  Consequently
\begin{equation}
x(t) = \frac{2}{b}\,\frac{t+k_1}{t^2+2k_1t+k_2}, \label{0.7a}
\end{equation}
where $k_1$ and $k_2$ are arbitrary constants.
The second of  \re{0.6} is zero if $\alpha b = 2 $.  In this case \re{0.5}
becomes
\begin{equation}
w\dddot{w} + 3\dot{w}\ddot{w} = 0 \label{0.8}
\end{equation}
which is a linear third order equation in the variable $w^2 $
with the same solution as \re{0.7} with the exception that the $w $ is replaced
by $w^2 $.  In either case the solution of \re{0.3} is the same when the
transformation \re{0.4} is applied.  It is amusing to note that, if \re{0.8} is
multiplied by the integrating factor $w^2 $ and integrated, we obtain
\begin{equation}
\ddot{w} =\frac{K}{w^3}, \label{0.8a}
\end{equation}
where
$K $ is the constant of integration, which is a particular instance of the
Ermakov-Pinney equation \cite{Ermakov, Pinney}.  Thus one can easily see the
relationship between the solutions of the two third order equations.  The
solution of the Ermakov-Pinney equation
\begin{equation}
\ddot{x} +\omega^2x = \frac{1}{x^3} \label{0.88}
\end{equation}
is given by
\begin{equation}
x = \sqrt{Au^2+2Buv+Cv^2},\quad AC-B^2 = W, \label{0.89}
\end{equation}
where $u$ and $v $ are two linearly independent solutions of
\begin{equation}
\ddot{x}+\omega^2x = 0 \label{0.90}
\end{equation}
and $W $ is the constant value of
their Wronskian.  Without the constraint on the constants of integration the
expression in the square root is simply the general solution of a third order
equation of maximal symmetry, the solution of which can be expressed in terms
of the three independent quadratic terms obtainable from the solution set of
the corresponding second order equation \cite{Moyo}.

We can transform the second order differential equation \re{0.3} into a
two-dimensional system of first order equations by a transformation which
preserves the symmetries in \re{0.1}.  We write
\begin{equation}
y = x\left(\frac{1}{\beta}\frac{\dot{x}}{x^2} - \frac{\alpha}{\beta}\right) \label{0.9}
\end{equation}
to obtain the system
\begin{eqnarray}
\dot{x} \z \alpha x^2 + \beta xy \label{0.10}\\
\dot{y} \z - \frac{1}{\beta}\left(\alpha + b\right)\left(2\alpha + b\right)x^2
- 3\left(\alpha + b\right)xy  - \beta y^2 \label{0.11}
\end{eqnarray}
in which the constants $\alpha $ and $\beta $ are arbitrary.

The generality of \re{0.3}, \re{0.10} and \re{0.11} is merely apparent.  In
\re{0.3} the parameter $b $ may be scaled out of the equation by the
transformation $bx \rightarrow x $ or, alternatively, by a rescaling of the
time variable.  In the case of equations \re{0.10} and \re{0.11} we can remove
the parameter $\beta $ by the rescaling of the new variable $y $ according to
the transformation $\beta y \rightarrow y $.  Consequently the analysis of
these equations can be performed without loss of generality on the second order
ordinary differential equation
\begin{equation}
\ddot{x} + 3x\dot{x} + x^3 = 0 \label{0.12}
\end{equation}
and the system of two first order equations
\begin{eqnarray}
\dot{x} \z a x^2 +  xy \label{0.13}\\
\dot{y} \z - \left(a + 1\right)\left(2a + 1\right)x^2 - 3\left(a + 1\right)xy
- y^2 \label{0.14}
\end{eqnarray}
obtained from \re{0.12} by the transformation
\begin{equation}
y =\frac{\dot{x}}{x^2} - ax \label{0.15}
\end{equation}
in
which we have replaced the parameter $\alpha $ by the parameter $a $ to aviod
confusion when we
use the former in the Painlev\'e analysis.  Note that $G_2 $ of the
original second order equation becomes
\begin{equation}
G_2 = -t\pt + x\px + y\py \label{0.16}
\end{equation}
for the two-dimensional system.

The two-dimensional system, \re{0.13} and \re{0.14}, is a particular instance
of a class of quadratic systems integrable by the factorisation method of
Adler-Kostant-Symes and its generalisation \cite{Golubchik,Golubchik2} which
have been recently studied from the point of view of their symmetry and
singularity properties \cite{Leach2}.

\section{Explicit integration of \sodes\ of the class}

We consider certain specific forms of \re{0.2}, \viz
\begin{equation}
\frac{\ddot{x}}{x^{3}}+f\(\frac{\dot{x}}{x^{2}}\)=0, \quad
x=x(t).\label{1}
\end{equation}
\vspace{3mm}

\subsection{A) $f(\dot{x}/x^{2})=2a\dot{x}/x^{2},$ $a$ real.}

Equation \re{1} becomes
\begin{equation}
\ddot{x}+2ax\dot{x}=0\label{2}
\end{equation}
and, since \re{2} can be written as
\begin{equation}
\frac{\d p}{\d x}=-2ax,\quad p(x)=\dot{x},\label{3}
\end{equation}
by carrying out the trivial integration of \re{3} we finally obtain for $x(t)$
\begin{equation}
x(t)=\sqrt{\frac{k}{a}}\left( \frac{k_{1}\e^{2t\sqrt{ak}}-1}{k_{1}\e^{2t\sqrt{%
ak}}+1}\right). \label{4}
\end{equation}

Equation \re{4} has a simple pole as singularity.
\vspace{3mm}

\subsection{B) $f(\dot{x}/x^{2})=2a\dot{x}/x^{2}+2b,$ $a$ and $b$ real.}

In this case \re{1} becomes
\begin{equation}
\ddot{x}+2ax\dot{x}+2bx^{3}=0.\label{5}
\end{equation}

Equation \re{5} appears in Kamke \cite{Kamke}[6.43, p 551],
but the method used there leads, even
for special values of $a$ and $b$, to integrals which are beyond any hope of
calculation. Therefore we resort to transforming \re{5} into a third-order
equation by means of the Riccati Ansatz \cite{Leach},
\begin{equation}
x=\alpha \frac{\dot{w}}{w},\quad w=w(t),\label{6}
\end{equation}
in which case \re{5} becomes
\begin{equation}
\frac{\dddot{w}}{w}+\frac{\ddot{w}\dot{w}}{w^{2}}(-3+
2a\alpha )+\frac{ \dot{w}{}^{3}}{w^{3}}(2-
2a\alpha +2b\alpha^{2})=0. \label{7}
\end{equation}

Choosing in \re{7}
\begin{equation}
\alpha =\frac{3}{2a },\quad b=\frac{2a^{2}}{9}\label{8}
\end{equation}
we finally obtain from \re{7} and \re{6} that the general solution of \re{5} is
\begin{equation}
x(t)=\frac{3}{a}\left( \frac{t+k_{1}}{t^{2}+2k_{1}t+k_{2}}\right) \label{9}%
\end{equation}
provided
\begin{equation}
b=\frac{2a^{2}}{9}, \label{10}
\end{equation}
where
$k_{1}$ and $k_{2}$ are constants of integration. Equation \re{9} has poles as
singularities. The results \re{9} and \re{10} for \re{5} are known \cite{Leach}
and are identical to \re{0.7a} for \re{0.3} provided that we identify the $a$ in
\re{5} with $3b/2$ in \re{0.3}.

In the following we drop the condition \re{10} and investigate the resulting
third-order equation \re{7}, \ie
\begin{equation}
w^{2}\dddot{w}+\dot{w}{}^{3}c=0,\quad
 c=\frac{9b}{2a^{2}}-1.\label{11}
\end{equation}

We reduce the order of \re{11} by using the standard substitution
\begin{equation}
\dot{w}=p,\quad p=p(w). \label{12}
\end{equation}
Hence, from \re{11} and \re{12}, we deduce that
\begin{equation}
w^{2}pp''+w^{2} p'{}^{2}+cp^{2}=0%
 \label{13}
\end{equation}
which is a linear \sode\ in the variable $p^2$ of Euler type.  Consequently
\re{13} has eight Lie point symmetries which is an increase from the two of the
original \sode.  The technique of increasing and decreasing the order of an
equation using different symmetries has been found often to be most useful
\cite{Barbara1,Barbara2}.

Equation \re{13} is treated in Kamke \cite{Kamke}[6.184, p 585]
and the result is
\begin{equation}
\left\{
\begin{array}{c}
p(w)=\left[ C_{1}w^{\rho _{1}}+C_{2}w^{\rho _{2}}\right] ^{1/2} \\  \\
\rho _{1}=\ha\(1-\sqrt{1-8c}\),\quad\rho _{2}=\ha\(1+\sqrt{1-8c}\),\quad
c\neq 1/8
\end{array}
\right\}. \label{14}
\end{equation}
Equations \re{12} and \re{14} yield
\begin{equation}
\int w^{-\rho _{1}/2}\lb C_{1}+C_{2}w^{\rho _{2}-\rho _{1}}\rb
^{-1/2}\d w=t+C, \label{15}
\end{equation}
where $C_{1}$, $C_{2}$ and $C$ are constants of integration.  By means of the
substitution
\begin{equation}
w^{\rho _{2}-\rho _{1}}=v^{\gamma },\quad\gamma =\frac{\rho _{2}-\rho _{1}%
}{1-\rho _{1}/2},
\quad\sqrt{1-8c}=\frac{3{ }\gamma }{4-\gamma }
\label{16}
\end{equation}
the integral, $I$, on the left hand side of \re{15} becomes
\begin{equation}
I{ =}\frac{2}{2-\rho _{1}}\int \left[ C_{1}+C_{2}v^{\gamma }\ \right]
^{-1/2}\d v.\label{17}
\end{equation}

Now, $I$ in \re{17} can be transformed into an integral of rational functions
\cite{GR}[pp 70 - 71] if $1/\gamma  =$ integer $=k$. The case $k=1$ leads, due
to \re{14} and \re{16}, to $\sqrt{1-8c}=1,$ \ie $c=0$, which in conjunction with
\re{10} and \re{11} produces the already known result, \re{9}. Let then $k\neq 1$ and
set
\begin{equation}
C_{1}+C_{2}v^{\gamma }=u^{2}.\label{18}
\end{equation}
By means of \re{18} the integral in \re{17} is written as
\begin{equation}
I=\frac{4}{C_{2}\gamma (2-\rho _{1})}\int \left( \frac{u^{2}-C_{1}}{C_{2}}%
\right)^{k-1}\d u. \label{19}
\end{equation}
In \re{19}\, $k-1= $ integer.  By virtue of \re{16} $0<\gamma <4$.  Thus from $%
1/\gamma $ = integer = $k$ we have $k=1,2,3,\ldots$. Let $k=2$, \ie $\gamma
=1/2 $ . Then by \re{11} and \re{16}
\begin{equation}
b=\frac{12a^{2}}{49}\label{20}
\end{equation}
 and from \re{19}, on inserting $\rho _{1}$ from \re{14} and \re{16} with $\gamma =1/2$,
and utilising \re{15}, we obtain
\begin{equation}
u^{3}-3C_{1}u-\frac{9(t + C)C_{2}^{2}}{14}=0,\quad u=u(t). \label{21}
\end{equation}

Owing to \re{6}, \re{16} and \re{18} we obtain, since $\gamma =1/2$,
\begin{equation}
x(t)=\frac{7u{ }\dot{u}}{a\(u^{2}-C_{1}\)}, \label{22}
\end{equation}
where $u(t)$ is given implicitly by \re{21}.

Equation \re{21} can be solved analytically for $u=u(t)$. To this end we introduce
the notations \cite{AS}
\begin{equation}
S_{1}=\left\{ \frac{9}{28}C_{2}^{2}(t+C)+\left[ -C_{1}^{3}+\frac{9^{2}}{%
28^{2}}C_{2}^{4}(t+C)^{2}\right]^{1/2}\right\}^{1/3}\label{23}
\end{equation}
and
\begin{equation}
S_{2}=\left\{ \frac{9}{28}C_{2}^{2}(t+C)-\left[ -C_{1}^{3}+\frac{9^{2}}{%
28^{2}}C_{2}^{4}(t+C)^{2}\right] ^{1/2}\right\} ^{1/3}.\label{24}
\end{equation}
Then, from \re{21},
\begin{equation}
\begin{array}{c}
u=S_{1}+S_{2}{ } \\
u=-\frac{1}{2}\left( S_{1}+S_{2}\right) +\frac{i\sqrt{3}}{2}\left(
S_{1}-S_{2}\right) \\
u=-\frac{1}{2}\left( S_{1}+S_{2}\right) -\frac{i\sqrt{3}}{2}\left(
S_{1}-S_{2}\right)
\end{array}
\label{25}
\end{equation}
and, after a modicum of calculation, from \re{22} - \re{25} we deduce,
considering firstly $u=S_{1}+S_{2}$ in \re{25},
\begin{equation}
x(t)=\frac{7Y(t)}{3a(u_{1}^{2}-1)\sqrt{-1+k^{2}(t+C)^{2}}},
\label{26}
\end{equation}
where $k$=$\frac{9}{28}C_{2}^{2}/C_{1}^{3/2}$ and $C$ are constants of
integration, and
\begin{equation}
\begin{array}{c}
s_{1}=\left\{ k(t+C)+\left[ -1+k^{2}(t+C)^{2}\right] ^{1/2}\right\} ^{1/3}
\\
s_{2}=\left\{ k(t+C)-\left[ -1+k^{2}(t+C)^{2}\right] ^{1/2}\right\} ^{1/3}
\\
u_{1}=s_{1}+s_{2}
\end{array}
\label{27}
\end{equation}
\begin{equation}
\begin{array}{c}
Y(t)= k\(s_1^2-s_2^2\right).
\end{array}
\label{28}
\end{equation}

It can be easily verified that $u_{1}^{2}\neq 1.$ Consequently, the exact
solution, \re{26} and \re{27}, of \re{5}, proviso \re{20}, possesses branch points as
singularities. We have restricted ourselves to one of the three possible
solutions $u(t)$ of \re{21} appearing in \re{25} in order to gain some insight
into the structure of the solutions of \re{5} without unduly complicating the
respective calculations, provided \re{20} holds. This clearly implies that in
the initial value problem consisting of \re{5} endowed with the initial
conditions $x(t_{0})=x_{0},\dot{x}(t_{0})=\dot{x}_{0}$, we take
into account initial conditions such that the unique solution $x(t),$
deriving of course from the basic equation \re{21}, can also be extracted from $%
u=S_{1}+S_{2}.$ The remaining possibilities for $u$ in \re{25} will most
probably lead, owing to the form of $S_{1}$ and $S_{2},$ to $x(t)$ possessing
branch points or essential singularities or both.

From \re{19} it is obvious that for $k=3,4,\ldots$ the $x(t)$, which now holds for
all $a$ and $b$ fulfilling $\sqrt{1-8c}=3/(4k-1),$ $c=(9b-2a^2)/(2a^{2}),$
is still being given by \re{22} while $u=u(t)$ is determined
implicitly through an algebraic equation of degree $2k-1$, the solution of
which
for $u=u(t)$ cannot be effected analytically. Nevertheless the
aforementioned $x(t)$ and $u=u(t)$ do represent an exact solution of \re{5}.
The remaining case for which $I$ in \re{17} can be transformed into an integral
of rational functions \cite{GR}[pp 70 - 71] is $1/\gamma $ $-1/2$ = integer = $k$%
. If $k=0$, then $\gamma =2$ and by \re{16} $\ b=0$, which is simply case A)
above. As $0<\gamma <4$, we have $k=1,2,3,\ldots$ and we set
\begin{equation}
C_{1}+C_{2}v^{\gamma }=u^{2}v^{\gamma }, \label{29}
\end{equation}
so that \re{15} - \re{17} yield
\begin{equation}
I{ = - }\frac{4C_{1}^{k}}{\gamma (2-\rho _{1})}\int \left[ u^{2}-C_{2}%
\right] ^{-(k+1)}\d u{ }=t+C, \label{30}
\end{equation}
while the equivalent to \re{22} is now
\begin{equation}
x(t)=\frac{3u{ }\dot{u}}{a\(C_{2}-u^{2}\)\sqrt{1-8c}}. \label{31}
\end{equation}

Clearly the integral in \re{30} is in principle calculable, but even for $k=1$
the resulting defining equation for $u=u(t)$ is not invertible. However, we
do have in implicit form, \ie \re{30} and \re{31}, an exact solution to \re{5}
provided $\sqrt{1-8c}=3/(4k+1),$ $c=(9b-2a^2)/(2a^{2}).$  We
do not think it is worthwhile to pursue this further, since the characteristic
features of the structure of the solutions to \re{5} are borne out by the cases
analytically calculated above. Note also that, either by setting one of the
two constants $C_{1}$ and $C_{2}$ in \re{15} equal to zero or directly from \re{5}, one
immediately obtains the particular solution to \re{5}, \viz
\begin{equation}
x(t)=\frac{\lambda }{t+C},\quad b\lambda ^{2}-a\lambda +1=0, \label{32}
\end{equation}
with one of the $\lambda $-values in \re{32} for $C=0$ corresponding to \re{9} and \re{10}
for $k_{1}=k_{2}=0.$  The solution \re{32} has a simple pole as singularity.

As can be expected,
nothing is gained by interchanging $\rho _{1}$ and $\rho _{2}$ in \re{14},
while $c=1/8$ in \re{14} leads to the appearance of $\log w$\ in the integral in
\re{15} thus rendering it not calculable.  This coincides with the result found
when the Painlev\'e test is applied to \re{5} \cite{Leach}.

We conclude the discussion of \re{5} by
noting that for $a=1/2$ and $b=-1/2$
\re{5} possesses the general solution \cite{Kamke}[6.32, p 548] \\    \\
\hspace*{\fill} $x(t)=C_{1}\dsp{\frac{\dot{\mathcal{P}}(C_{1}t+C_{2};0,1)}{{\mathcal{P}}%
(C_{1}t+C_{2};0,1)}}$\hfill$ {\mbox{\rm (\ref{32}$^1$)}}$         \\  \\
${\mathcal P}(C_1t+C_2;0,1)$\ being the Weierstrass elliptic function \cite{GR}[p
917], where the prime denotes differentiation with respect to $C_1t+C_2$,
and $C_1$ and $C_2$\ are constants of integration. By using the series expansion of $%
{\mathcal P}(C_1t+C_2;0,1)$\ it is seen that $x(t)$\
in (\ref{32}$^1$) has a simple pole at $%
C_1t+C_2=0.$

Finally, if in \re{5} we set\\  \\
\hspace*{\fill} $x = u^{-1/2},$ \hfill$\mbox{\rm (\re{32}$^{2}$)}$\\  \\
\re{5} becomes\\  \\
\hspace*{\fill}$\ddot{u}-\tha\dot{u}^2u^{-1}+2a\dot{u}u^{-1/2}-4b=0.$
 \hfill$\mbox{\rm
(\ref{32}$^3$)}$\\  \\
Then due to\\  \\
\hspace*{\fill}$\dot{u} = p,\quad p = p(u)$ \hfill$\mbox{\rm (\ref{32}$^4$)}$\\
\\
we deduce from (\ref{32}$^{^{3}}$)\\   \\
\hspace*{\fill}$up\dot{p}-\tha p^2+2apu^{1/2}-4bu = 0.$
\hfill$\mbox{\rm (\ref{32}$^5$}$\\  \\
Equation (\ref{32}$^5$) \ is an Abel's equation of the second kind and can be
treated according to Kamke \cite{Kamke}[pp 25-28]. The final result obtained after
some calculations is\\ \\
\hspace*{\fill}$\dsp{\begin{array}{c}
(-2a)^{-1/2}\int z^{-1-r_{1}/2}\(C_{1} +C_{2}z^{r_{2}-r_{1}}\)^{-1/2}\d
z=t+C_{3} \\
u=-2a\(C_{1}z^{r_{1}}+C_{2}z^{r_{2}}%
\)^{-1},\quad  z=z(t) \\
r_{1} =\(1-\sqrt{1-4b/a^{2}}\)/2,\quad r%
_{2} =\(1+\sqrt{1-4b/a^{2}}\)/2.
\end{array}          }$
\hfill\mbox{\rm (\ref{32}$^{6}$}\\ \\
There are two independent constants in (\ref{32}$^{6}$) due to (\ref{32}$^{3}$).
Equations (\ref{32}$^{6}$) are essentially equivalent to \re{16} - \re{19}
 and one
obtains the same results as previously. Therefore transformation (\ref{32}$^{2}$)
does not lead to any new results other than indicating another way to
obtain the results already known. \ If instead of (\ref{32}$^{2}$)
we consider $%
x=-u^{-1/2},$ equivalent results are\ extracted.
\vspace{3mm}

\subsection{C) $f(\dot{x}/x^{2})=a(\dot{x}/x^{2})^{2}+b%
\dot{x}/x^{2}+c,$ with $a$, $b$ and $c$ real and $a\neq 0$.}

Now \re{1} becomes
\begin{equation}
x\ddot{x}+a(\dot{x})^{2}+bx^{2}\dot{x}+cx^{4}=0%
\label{33}
\end{equation}

Equation \re{33} appears in Kamke \cite{Kamke}[6.130, p 574] and we employ the
method described there at the end of this paragraph. In the following
by means of  \re{6} we obtain from \re{33} by choosing
\begin{equation}
\alpha =\frac{2a+3}{b}\label{34}
\end{equation}
\begin{equation}
\begin{array}{c}
\dddot{w}+a\dsp{\frac{ \ddot{w}{}^{2}}{\dot{w}}}+
d\dsp{\frac{ \dot{w}{}^{3}}{w^{2}}}=0 \\
d=a+2+c\left(\dsp{ \frac{2a+3}{b}}\right)^{2}-(2a+3).
\end{array}
\label{35}
\end{equation}

We firstly assume
\begin{equation}
d=0\label{36}
\end{equation}
which implies one relation between the coefficients $a$, $b$ and $c.$ Then
\re{35} gives
\begin{equation}
\begin{array}{c}
\dot{w}=y \\
\ddot{y}+a\dsp{\frac{\dot{ y}{}^{2}}{{y}}}=0
\end{array}
\label{37}
\end{equation}
the second of which is linear in $y^{a+1}$.

Equation \re{37} is treated in Kamke \cite{Kamke}[6.128, p 574]. We obtain, by taking
into account \re{6} after some straightforward calculation,
\begin{equation}
x(t)=\frac{\left( \dsp{\frac{2a+3}{b}}\right)
\left(\dsp{ \frac{a+2}{a+1}}\right) }{%
k_{2}(t+k_{1})^{-1/(a+1)}+(t+k_{1})\ }{ ,}\quad a\neq -1 \label{38}
\end{equation}
as the solution to \re{33} if \re{36} is valid, $k_{1}$ and $k_{2}$ being
constants of integration.
If $1/(a+1)=-k$, $k=1,2,\ldots$ or $1/(a+1)=k$, $k=1,2,\ldots$,
the $x(t)$
in \re{38} possesses poles of order $k$ at the most or poles of order $k+1$ at
the most as singularities respectively. For all other values of $1/(a+1)$ \re{38} may
possess all kinds of singularities, for example a pole and an essential
singularity if $\ -1-1/(a+1)\neq \mbox{\rm integer }>0$, an essential singularity
and a branch point if $1/(a+1)+1\neq \mbox{\rm integer }>0$ and so on. In the case $%
a=-1$ \re{38} becomes
\begin{equation}
x(t)=\frac{k_{1}e^{k_{1}t}}{b\(k_{2}+e^{k_{1}t}\)},\quad a = -1%
\label{39}
\end{equation}
which has a simple pole as singularity.

In what follows we discard condition \re{36} and examine the full equation
\re{35}. Then
\begin{equation}
\begin{array}{c}
\dot{w}=p,{ }p=p(w) \\
w^{2}pp^{^{\prime \prime }}+w^{2} p^{^{\prime }}{}^{2}(a+1)+dp^{2}=0
\end{array}
\label{40}
\end{equation}
which is an equation of Euler type linear in the variable $w^{a+2}$ provided $a
\neq -2$.
Equation \re{40} is precisely of the form \re{13} treated in case B) above. Before
proceeding further we remark that the case $a=-2$ in \re{40} mentioned in
Kamke \cite{Kamke}[6.184, p 585] leads to $p(w)=kw^{d}\exp(C_{1}w)$ $=\dot{w}$ and
the resulting relation cannot be inverted for $w=w(t).$  Equation \re{40} is
solved and, as in \re{14}, we obtain
\begin{equation}
\left\{
\begin{array}{c}
p(w)=\left[ C_{1}w^{\rho _{1}}+C_{2}w^{\rho _{2}}\right] ^{1/(a+2)} \\  \\
\rho _{1}=\ha\(1-\sqrt{1-4(a+2)d}\),\quad\rho _{2}=\ha\(1+\sqrt{1-4(a+2)d}\)%
,\quad 4(a+2)d{ }\neq 1
\end{array}
\right\}. \label{41}
\end{equation}
Consequently due to \re{40}
\begin{equation}
I=\int w^{-\rho _{1}/(a+2)}{ }\left[ C_{1}+C_{2}w^{\rho _{2}-\rho _{1}}%
\right] ^{-1/(a+2)}\d w=t+C. \label{42}
\end{equation}
If we set
\begin{equation}
w^{\rho _{2}-\rho _{1}}=v^{\gamma },\quad\gamma =\frac{(\rho _{2}-\rho
_{1})(a+2)}{a+2-\rho _{1}},\quad\sqrt{1-4(a+2)d}=\frac{\gamma (a+3/2)}{%
a+2-\gamma /2},\label{43}
\end{equation}
the integral, $I$, in \re{42} takes the form
\begin{equation}
I{ =}\frac{a+2}{a+2-\rho_{1}}\int \left[ C_{1}+C_{2}^{\gamma }\ %
\right] ^{-1/(a+2)}\d v =t+C \label{44}
\end{equation}
and, by virtue of the substitution
\begin{equation}
C_{1}+C_{2}v^{\gamma }=u^{a+2},\label{45}
\end{equation}
\re{44} yields
\begin{equation}
I=\frac{a+2}{C_{2}\sqrt{1-4(a+2)d}}\int u^{a}\left( \frac{u^{2}-C_{1}}{C_{2}}%
\right) ^{r-1}\d u{ }=t+C{,\quad }1/\gamma =r. \label{46}
\end{equation}
The only possibility existing for $I$ in \re{46}\ to be computable is when both
$a$ and $r$ are rational and in particular \cite{GR}
\begin{description}
\item
i) $r =$ integer
\item
ii) $(a+1)/2 =$ integer
\item
iii) $(a+1)/2 + r =$ integer.
\end{description}

If $r=1$, we arrive at the known cases $a=-2$ or\ $d=0$. The next simplest
case is $a=1$ and $r=2$, \ie $\gamma =1/2$. Then from \re{35}, \re{43} and \re{46}
\begin{equation}
\begin{array}{c}
I=\dsp{\frac{3}{C_{2}\sqrt{1-12d}}\int u\left( \frac{u^{2}-C_{1}}{C_{2}}\right)
^{\ }\d u{ }}=t+C{ } \\
d= \dsp{\frac{8}{121}},\quad d=-2+25c/b^{2},\quad a=1.
\end{array}
\label{47}
\end{equation}
From \re{6}, \re{34}, \re{43}, \re{45} and \re{47} with $a=1$ and
$\gamma =1/2$ we deduce that
\begin{equation}
u^{4}-2u^{2}C_{1}-\frac{20}{33}C_{2}^{2}(t+C)=0\label{48}
\end{equation}
\begin{equation}
x(t)=\frac{11}{b}\left( \frac{3u^{2}\dot{u}}{u^{3}-C_{1}}\right).
\label{49}
\end{equation}
Relations \re{48} and \re{49} constitute an exact solution of \re{33} provided
the conditions \re{47}
are satisfied. Equation \re{48} can be solved for $u(t)$ and along the same lines as
in Case B) above we arrive at a singularity pattern similar to that of \re{22},
\ie branch points and essential singularities.

We think that there is no essence in pursuing this further by examining
other cases in i) - iii) above since we have already obtained a very general
singularity behaviour of the solutions to \re{33} through \re{48} and \re{49}.

Now, in contrast to case B) above, some further exact results can be
extracted for \re{33} by following Kamke \cite{Kamke}[{\it loc cit}]
and we obtain for \re{33}
\be
\begin{array}{c}
p=\dot{x},\quad p=p(x)=x^2u_1,\quad u_1=u_1(t_1),\quad t_1=\ln x \\
u_1^{^{\prime }}u_1+u_1^2(a+2)+bu_1+c=0
\end{array}
\label{50c}
\ee
whereas, after a straightforward calculation,
\be
\begin{array}{c}
a'=\dsp{\frac{ b}{a+2}},\quad b'=\dsp{\frac{ c}{a+2}} \\
\log \left[ \left( u_1^2+a'u_1+b'\)^{1/2}
\right] -\log \left[ \left(\dsp{ \frac{\sqrt{a'{}^2-4b'}%
-a'-2u_1}{a'+2u_1+\sqrt{a'{}^2-4b'}}}\right)
^{a'/[2\sqrt{a'{}^2-4b'}]}\right]\n\\
 = \log \left[ \left( C_1/x\right) ^{a+2}\right],\quad C_1=cons\tan t.
\end{array}\label{51c}
\ee

For \re{51c} to be of any use at all the quantity $a'/[2\sqrt{%
a'{}^2-4b'}]$ has to be an integer and we consider
in the following the simple case
\be
\ha a'= \sqrt{a'{}^2-4b'}\label{52c}%
\ee
and, therefore, \re{51c} finally yields
\be
\begin{array}{c}
u^3-uw+\ha a'=0 \\
u=u_1+\tqr a',\quad w=\left( C_1/x\right) ^{2a+4}.
\end{array}
\label{53c}
\ee

Along the lines of case B) above and by applying \re{50c} we deduce
\be
x'=x^2\left\{
\begin{array}{c}
-\frac{3a^{^{\prime }}}4+\left[ -\frac{a^{^{\prime }}}4+\left( -\frac{%
C_1^{6a+12}}{9x^{6a+12}}+\frac{(a^{^{\prime }})^2}{16}\right) ^{1/2}\right]
^{1/3} \\
+\left[ -\frac{a^{^{\prime }}}4-\left( -\frac{C_1^{6a+12}}{9x^{6a+12}}+\frac{%
(a^{^{\prime }})^2}{16}\right) ^{1/2}\right] ^{1/3}.
\end{array}
\right\} \label{54c}
\ee
On integrating \re{54c} we obtain
\be
\begin{array}{c}
-\int \left\{
\begin{array}{c}
-\frac{3a^{^{\prime }}}4+\left[ -\frac{a^{^{\prime }}}4+\left( lz^{6a+12}+%
\frac{(a^{^{\prime }})^2}{16}\ \right) ^{1/2}\right] ^{1/3} \\
+\left[ -\frac{a^{^{\prime }}}4\ -\left( lz^{6a+12}+\frac{(a^{^{\prime }})^2}{%
16}\ \right) ^{1/2}\right] ^{1/3}
\end{array}
\right\} ^{-1}dz=t+C_2 \\
z=1/x,\quad l=-C_1^{6a+12}/9.
\end{array}
\label{55c}
\ee

The integral in \re{55c} can be brought into a form more amenable to
treatment by setting
\be
\left[ -\frac{a^{^{\prime }}}4+\left( lz^{6a+12}+\frac{(a^{^{\prime }})^2}{16%
}\ \right) ^{1/2}\right] ^{1/3}+\left[ -\frac{a^{^{\prime }}}4-\left(
lz^{6a+12}+\frac{(a^{^{\prime }})^2}{16}\ \right) ^{1/2}\right] ^{1/3}=y%
\label{56c}
\ee
whence
\be
\begin{array}{c}
-(3l^{1/3})^{-\frac 1{2a+4}}\left( \frac 1{2a+4}\right) \left\{
\begin{array}{c}
\int \left[ (-y^3-\frac{a^{^{\prime }}}2)^{-\frac{2a+3}{2a+4}}y^{\frac{4a+7}{%
2a+4}}(-3)^{^{-\frac 1{2a+4}}}-(-y^3-\frac{a^{^{\prime }}}2)^{\frac
1{2a+4}}y^{^{-\frac{2a+5}{2a+4}}}\right] \frac{dy}{y-\frac{3a^{^{\prime }}}4}
\\
\
\end{array}
\right\} = \\
t+C_2.
\end{array}
\label{57c}
\ee

If we reflect a little on \re{57c} we conclude that only for specific $%
a $-values can we attempt to carry out the integration in \re{57c}. For
example, if $a=-3/2$, then \re{51c}, \re{52c} and \re{55c} - \re{57c} yield
\be
\begin{array}{c}
\frac{2y}3+\frac 2{3y}+\log \left[ \left| \frac{(y-3b/2)^{b+4/9b}}{y^{4/9b}}%
\right| \right] =k(t+C_2) \\
x(t)=\frac y{k(y^3+b)},\quad y=y(t)
\end{array}
\label{58c}
\ee
as an exact solution to (2.33), where $C_2$ and $k$ are constants, provided
\be
a=-3/2,\quad c=3b^2/8,\quad b>0\label{59c}.
\ee
We remark that if we replace \re{52c} by $\ha a' =-\sqrt{%
a'{}^2-4b'}$ we again arrive at \re{58c} - \re{59c}, the
only difference being that \re{59c} also holds for $b<0$.\ From \re{58c} it also
follows that $x(t)$ has a singularity at $t=t_s,$\ $t_s$\ being calculated
from \re{58c} by setting $y=(-b)^{1/3}.$\

Evidently we can find other $a$-values, for instance $a=-11/6,$ for which
the integration in \re{57c} is possible and we would arrive at formul\ae\ similar
to \re{58c}. Likewise we may take into account other possibilities in place of
\re{52c}. Nevertheless there is little point in carrying this investigation
further since it is almost certain that the resulting relations equivalent
to \re{58c} will not be invertible for the auxilliary function $y(t)$.

\section{Painlev\'e analysis of the second order\newline equation}

We determine the leading order behaviour of \re{0.12} by means of the
substitution $x = \alpha\tau ^p $ and find that the value of $p $ is $- 1 $ in
accordance with the coefficients of the rescaling symmetry, $G_2$, and
that the coefficient, $\alpha$, is a solution of the equation
\begin{equation}
\alpha^2 - 3\alpha + 2 = 0, \label{1.1}
\end{equation}
\ie $\alpha = 1,2 $.  To find the
resonances we make the substitution $x = \alpha x^{-1} + \beta x^{r-1}$.  The
condition that $\beta$ be arbitrary is
\begin{equation}
r^2 + 3(\alpha - 1)r + 3\alpha - 4 = 0, \label{1.2}
\end{equation}
\ie $ r=-1,4-3\alpha$.  With the
two possible values of $\alpha$ we have $r=-1,1$ for $\alpha = 1$ and $r=-1,-2$
for $\alpha=2$.  The former gives the standard Right Painlev\'e Series,
representing a Laurent series in the neighbourhood of the moveable singularity,
and the latter a Left Painlev\'e Series, representing a Laurent expansion valid
over the exterior of a disc surrounding the moveable singularity \cite{Feix}.

In terms of $\tau=t-t_0$ the explicit expressions for the series are
\begin{equation}
x(\tau) = \frac{1}{\tau}\left(1+ a_1\tau - a_1^2\tau^2+ a_1^3\tau^3 -
a_1^4\tau^4+\ldots\right) \label{1.3}
\end{equation}
for the Right Painlev\'e Series and
\begin{equation}
x(t) = 2t^{- 1} + a_1t^{- 2} + a_2t^{- 3} -
\ha a_1\left(a_1^2-3a_2\right)t^{-4} + \ha\left(a_1^2 -
a_2\right)^2t^{-5} \label{1.4}
\end{equation}
for the Left
Painlev\'e Series.  For the latter we have written the expansion in terms of $t
$ instead of $\tau $ because of the asymptotic nature of the expansion
\cite{Feix}.

We can reconcile the two Laurent series, \re{1.3}
and \re{1.4}, with the explicit solution, \re{0.7a} with $b=1$, given in the
previous section.  In the case of the Right Painlev\'e Series we have
\begin{eqnarray}
x(t) \z \frac{1}{\tau}\lb
2-\(1-a_1\tau+a_1^2\tau^2-a_1^3\tau^3-a_1^4\tau^4+\dots\)\rb \n\\
\z \frac{1}{\tau}\lb 2 - \frac{1}{1+a_1\tau}\rb \n\\
\z \frac{1+2a_1\tau}{\tau(1+2a_1\tau)} \label{1.13a}
\end{eqnarray}
which is just \re{0.7a} written slightly differently.  If we now expand this as
\begin{eqnarray}
x(t) \z \frac{2}{t}\frac{\lb 1-\dsp{\frac{1}{t}}\(t_0-
\dsp{\frac{1}{2a_1}}\)\rb}{\(1-\dsp{\frac{t_0}{t}}\)\lb 1-
\dsp{\frac{1}{t}}\(t_0-\dsp{\frac{1}{a_1}}\)\rb} \n\\
\z 2t^{-1}
\(2t_0-\frac{1}{a_1}\)t^{-2}+\(2t_0^2-\frac{2t_0}{a_1}+\frac{1}{a_1^2}\)t^{-3}\n\\
&&+\(2t_0^3-\frac{3t_0^2}{a_1}+\frac{3t_0}{a_1^2}-\frac{1}{a_1^3}\)t^{-4}+\ldots,
\label{1.14a}
\end{eqnarray}
we can verify that the expansion is identical to that in \re{1.14} when we
rename the coefficients of $t^{-2}$ and $t^{-3}$ in \re{1.14a}.

\section{Painlev\'e analysis of the two-dimensional\newline system}

To determine the leading order behaviour of the system, \re{0.13} and
\re{0.14}, we make the substitutions
\begin{equation} x = \alpha \tau^{p},\quad y = \beta \tau^q \label{1.5}
\end{equation}
to obtain
\begin{eqnarray}
\alpha p \tau^{p- 1} \z a\alpha^2\tau^{2p} +\alpha\beta\tau^{p+q} \n\\
\beta q\tau^{q- 1} \z - (1+a) (1+ 2a)\alpha^2\tau^{2p} - 3(a+1)
\alpha\beta\tau^{p+q} -\beta^2\tau^{2q} \label{1.6}
\end{eqnarray}
so that the powers to be compared are
\begin{equation}
\begin{array}{rrrr}
- 1 &p &q&\\
q- 1 & 2p &p+q & 2q
\end{array} \label{1.7}
\end{equation}
in which we have
removed a common $p $ from the first line.  If the first term in the right side
of \re{1.6} were zero, we would also be able to subtract a common $q $ from the
second line.  Evidently there are three possible systems which should be
analysed separately.

\subsection{The general case}

In general the product $(1 +a) (1+ 2a)\neq 0 $.  In this case we can have two
possible types of leading order behaviour.  The first is the generic one for
which $p = - 1 =q $.  The second is $q = - 1,p > - 1 $.

We commence our analysis with the first case.  The two coefficients $\alpha $
and $\beta $ are found from the solution of
\begin{eqnarray}
-\alpha \z a\alpha^2+\alpha\beta \n\\
-\beta \z - (1+a) (1+ 2a)\alpha^2-3(1+a)\alpha\beta -\beta^2 \label{1.8}
\end{eqnarray}
and we obtain $\alpha =1,2 $ and $\beta = - (1+a), - (1+ 2a) $ respectively.

To determine the resonances we make the substitution
\begin{eqnarray}
x \z \alpha \tau^{- 1} +\mu\tau^{r- 1} \n\\
y \z \beta \tau^{- 1} + \nu\tau^{r- 1}  \label{1.9}
\end{eqnarray}
which allows both sets of coefficients to be treated at once.  The condition
that the coefficients $\mu $ and $\nu $ be arbitrary is given by
\begin{equation}
\left| \begin{array}{cc} r-a\alpha & -\alpha\\ (1+a) (a\alpha
+ 2\alpha - 3) &r- (3-3\alpha -a\alpha) \end{array} \right| = 0 \label{1.10}
\end{equation}
with the solution $r = - 1,4- 3\alpha $ so that in the case
$\alpha = 1 $ the resonances are at $r = - 1,1 $, indicating a Right Painlev\'e
series, and in the case $\alpha = 2 $ $r = - 1, - 2 $, indicating a Left
Painlev\'e series.  Note that the resonances are independent of the value of $a $.
The coefficients at the resonances are given by
\begin{equation}
\left(
\begin{array}{l}\mu\\\nu\end{array} \right) = \left(
\begin{array}{l}1\\1-a\end{array} \right) \sigma\quad\mbox{\rm and}\quad \left( \begin{array}{l}
-1\\1+a\end{array} \right) \sigma, \label{1.11}
\end{equation}
where $\sigma $
is a parameter, for the cases $\alpha = 1 $ and $\alpha = 2 $ respectively.

The other possible leading order behaviour is given by $q = - 1 $ and $p > - 1
$ which means that the series for $x $ starts at the constant term and so the
standard Painlev\'e analysis cannot be used.  We write
\begin{equation}
x = a_i\tau^{i} \qquad y = b_i\tau^{i-1} \label{1.12}
\end{equation}
in which the summation convention is adopted and
the counter $i $ commences at zero.  We substitute \re{1.12} into the system
\re{0.13} and \re{0.14} and obtain
\begin{eqnarray}
ia_i\tau^{i- 1} \z aa_ia_j\tau^{i+j} +a_ib_j\tau^{i+j- 1} \n\\
&&\label{1.13}\\
(i- 1)b_i\tau^{i- 2} \z - (1+a )( 1+ 2a)a_ia_j\tau^{i+j} -
3 (1+a)a_ib_j\tau^{i+j- 1} -b_ib_j\tau^{i+j- 2}. \n\\
\end{eqnarray}
We equate the coefficients of like powers of $\tau
$ to zero and obtain
\begin{equation}
\begin{array}{ll}
a_0 = 0 &b_0 = 1\\
a_1 = a_1 &b_1 = 0\\
a_2 = 0 &b_2 = 3 (1+a)a_1\\
a_3 = \ha (3+ 4a)a_1^2\quad &b_3 = 0\\
a_4 = 0 &b_4 = -\ofi (1+a) (19+ 20a)a_1^2.
\end{array} \label{1.14}
\end{equation}
It is a simple matter to verify that the first few terms of the
series for $x $ give the first few terms of the series for $y $ when
substituted into \re{0.15}.

\subsection{The case $(1+a) (1+ 2a) = 0 $}

As there are two roots to this quadratic equation, we have to consider two
subcases.

\subsubsection{$1+ 2a = 0 $}

The two-dimensional system, \re{0.13} and \re{0.14}, is now
\begin{eqnarray}
\dot{x} \z -\ha x^2+xy \n\\
\dot{y} \z -\tha xy -y^2. \label{1.15}
\end{eqnarray}
For the case when all terms are dominant, \viz $p =q = - 1 $, the
coefficients of the leading order terms are found from the solution of the
system
\begin{eqnarray}
-\alpha = -\ha\alpha^2 +\alpha\beta \n\\
-\beta = -\tha\alpha\beta -\beta^2 \label{1.16}
\end{eqnarray}
and are $\alpha = 1 $
and $\beta = -\ha $.  The resonances are given by $r =\pm 1 $ and so we
obtain a Right Painlev\'e series.  The arbitrary vector at the resonance $r = 1 $
is
\begin{equation}
\left( \begin{array}{l}\mu\\\nu\end{array} \right) =
\left( \begin{array}{l}2\\3\end{array} \right) \sigma.  \label{1.17}
\end{equation}

In addition to all terms contributing to the dominant behaviour there is the
possibility that just some of the terms can be dominant.  Such terms are termed
subdominant. These terms must have the same symmetry which will be more
restrictive than the rescaling symmetry of the system as a whole and this
property is termed subselfsimilarity.
The first possible subdominant behaviour is given by $p = - 1,q > - 1 $.  As
the series for $y $ does not contain a singularity, we must, as we did above,
substitute a full series into the system \re{1.15}.  We assume that
\begin{equation} x = a_i\tau^{i- 1} \quad y = b_i\tau^i \label{1.18}
\end{equation} with the same conventions as before.  We obtain the coefficients
\begin{equation}
\begin{array}{ll} a_0 = 2 &b_0 = 0\\ a_1 = 0 &b_1 = 0\\ a_2 =
0 &b_2 = 0\\ a_3 = 0 &b_3 = 0\\ a_4 = 0 &b_4 = 0 \end{array} \label{1.18a}
\end{equation}
so that we only obtain an isolated solution.

There is another possibility and that is that the subdominant behaviour
indicates the existence of an asymptotic series, \ie we have $p = - 1,q < - 1
$.  We write
\begin{equation}
x = a_it^{-i- 1} \quad y = b_it^{-i- 2} \label{1.19}
\end{equation}
with the counter beginning at zero.  Note that we
do not expand in terms of $\tau $ since the expansion is asymptotic.  We obtain
the following set of coefficients
\begin{equation}
\begin{array}{ll}
a_0 = 2&b_0 = 0\\
a_1 = a_1 &b_1 = b_1\\
a_2 = \ha a_1^2-2b_1 &b_2 = \tha a_1b_1\\
a_3= \oqr\(a_1^2-8b_1\)a_1 &b_3 = \tha\(a_1^2-b_1\)b_1
\end{array} \label{1.20}
\end{equation}
and, if the first few terms of the series for $x $ are
substituted into \re{0.15} with this value of $a $, we obtain verification of
the first few terms of the series for $y $.

\subsubsection{$1+a = 0 $}

The system is now
\begin{eqnarray}
\dot{x} \z - x^2+xy \n\\
\dot{y} \z -y^2. \label{1.21}
\end{eqnarray}
In the case that all terms are dominant the coefficients of the leading order
terms are given by
\begin{eqnarray}
-\alpha \z -\alpha^2 +\alpha\beta \n\\
-\beta \z  -\beta^2 \label{1.22}
\end{eqnarray}
and we find that $\alpha = 2
$ and $\beta = 1 $.  The resonances occur at $r = - 1, - 2 $ and consequently
we have a Left Painlev\'e series.

One of the possible patterns of subdominant behaviour is when $q = - 1,p > - 1
$.  We write
\begin{equation}
x = a_i\tau^{i} \quad y = b_i\tau^{i-1} \label{1.23}
\end{equation}
with the usual convention and find that we obtain the coefficients
\begin{equation}
\begin{array}{ll}
a_0 = 0 &b_0 = 1\\
a_1 = 0&b_1 = 0\\
a_2 = 0 &b_2 = 0\\
a_3 = 0 &b_3 = b_1\\
a_4 = 0 &b_4 = 0\\
a_5 = 0&b_5 = 0\\
a_6 = 0 &b_6 = \oth b_1^2
\end{array} \label{1.24}
\end{equation}
which is very strange since we are generating a solution for $y $ although it
is defined in terms of a zero function.

If we try the pattern $q = - 1,p < - 1 $, we find that the first few
coefficients are
\begin{equation}
\begin{array}{ll}
a_0 = 0 &b_0 = 1\\
a_1 = 0&b_1 = b_1\\
a_2 = 0 &b_2 = b_1^2\\
a_3 = 0 &b_3 = b_1^3\\
a_4 = 0 &b_4 =b_1^4
\end{array} \label{1.25}
\end{equation}
from which it is evident that we have the solution to the system to be
\begin{eqnarray}
x \z 0\n\\
y \z \frac{1}{t} +\frac{b_1}{t^2}+\frac{b_1^2}{t^3}+\frac{b_1^3}{t^4}
+\ldots\n\\ \z \frac{1}{t-b_1} \label{1.26}
\end{eqnarray}
which is a one
parameter solution of more than slightly suspect appearance.

\section{Comment}

We can make several observations of the results of these calculations.
\begin{enumerate}
\item
There is a close relationship between the possession of the Painlev\'e Property
and the ease of explicit integration of this class of equations.  Since they
are all second order equations with two symmetries, they are reducible to
quadratures and so integrable in the sense of Lie.  However, as we have seen in
Section 2, integrability in the sense of Painlev\'e is a much rarer occurrence.
We recall that integrability in the sense of Painlev\'e imposes stronger
requirements upon the solution since the function must be meromorphic.
\item
The properties of the two-dimensional system are
independent of the value of the parameter $a $ in the definition of the new
variable $y $, except for the two particular values of $a = - 1, -\ha $.  It is
interesting to note that the quadratic equation which determines these values
of $a $ is the adjoint of the equation which determines the value of the
coefficient, $\alpha $, of the leading order term in the analysis of the
original second order equation, \re{1.1}.
\item
In addition to the possibility
of the existence of the Left Painlev\'e Series and the Right Painlev\'e Series we
have seen the possibility of the existence of an asymptotic expansion which
really has nothing to do with the Painlev\'e analysis even though it indicates
singular behaviour.
\item
What is most disturbing is the existence of
solutions which are apparently spurious.  It remains an open question as to the correctness of the
proper interpretation of these solutions.
\end{enumerate}

\section*{Acknowledgements}
PGLL thanks the Dean of the School of Sciences, University of the Aegean,
Professor G P Flessas, and the Director of GEODYSYC, Dr S Cotsakis, for their
kind hospitality while this work was undertaken and the National Research
Foundation of South Africa and the University of Natal for their continuing
support.

\end{document}